\documentclass[apj,twocolumn]{openjournal}
\usepackage[dvipsnames]{xcolor}
\usepackage[T1]{fontenc}
\usepackage[utf8]{inputenc}
\usepackage[english]{babel}
\usepackage{natbib}

\usepackage{hyperref}
\hypersetup{
    unicode, 
    colorlinks=true,
    linkcolor=linkcolor,
    citecolor=linkcolor,
    filecolor=linkcolor,
    urlcolor=linkcolor,
}
\usepackage{color,colortbl}
\definecolor{linkcolor}{rgb}{0.0,0.3,0.5}
\DeclareGraphicsExtensions{.bmp,.png,.jpg,.pdf}
\usepackage{orcidlink}

\usepackage{savesym}
\savesymbol{tablenum}
\usepackage{siunitx}
\restoresymbol{SIX}{tablenum}
\usepackage{float}
\usepackage{tabularray}
\usepackage{booktabs}
\usepackage{bm}

\usepackage{microtype}

\UseTblrLibrary{booktabs}

\DeclareSIUnit{\erg}{erg}

\newcommand{\sfrac}[2]{\mathchoice%
  {\kern0em\raise.5ex\hbox{\the\scriptfont0 #1}\kern-.15em/
    \kern-.15em\lower.25ex\hbox{\the\scriptfont0 #2}}
  {\kern0em\raise.5ex\hbox{\the\scriptfont0 #1}\kern-.15em/
    \kern-.15em\lower.25ex\hbox{\the\scriptfont0 #2}}
  {\kern0em\raise.5ex\hbox{\the\scriptscriptfont0 #1}\kern-.2em/
    \kern-.15em\lower.25ex\hbox{\the\scriptscriptfont0 #2}} {#1\!/#2}}

\newcommand{\castro}{{\sf Castro}}

\newcommand{\amrex}{{\sf AMReX}}
\newcommand{\pynucastro}{{\sf pynucastro}}

\ExplSyntaxOn

\cs_set:Npn \__isot_helper:ee #1#2 { \ensuremath{{}^{#2}}\text{\text_titlecase_first:n {#1}} }

\seq_new:N \l_isot_parts
\regex_const:Nn \c_isot_regex { ^ ([A-Za-z]+) (\d+) $ }
\msg_new:nnnn { isot } { no-match } { invalid~isotope~format } { \c_backslash_str~isot~expects~a~single~argument~of~the~form~<element><mass~number>. }
\NewDocumentCommand \isot {m} {
  \tl_set:Nn \l_tmpa_tl {#1}
  \regex_extract_once:NnNTF \c_isot_regex {#1} \l_isot_parts
    { % pop the entire match element from the front and discard it
      \seq_pop_left:NN \l_isot_parts \l_tmpa_tl
      \__isot_helper:ee {\seq_item:Nn \l_isot_parts 1} {\seq_item:Nn \l_isot_parts 2}
    }
    { \msg_error:nn { isot } { no-match }
      {#1}
    }
}

\cs_set:Npn \rxnseq_format_part:nn #1#2 {
  \str_case:nnF {#2} {
    {a}  { $\alpha$ }
    {aa} { $\alpha\alpha$ }
    {g}  { $\gamma$ }
    {e-} { $\beta^{-}$ }
    {e+} { $\beta^{+}$ }
    {nu} { $\nu$ }
    {,}  { ,
      \str_if_eq:eeF {\seq_item:Nn \l_rxnseq_parts {\int_eval:n { (#1) + 1 }}} {)} { \nobreakspace }
    }
    {~}  {}
    {)}  { ) \allowbreak } % allow line breaks after right parentheses (after fisker:2008)
  }
  { % didn't match a case, try parsing as an isotope
    \regex_extract_once:NnNTF \c_isot_regex {#2} \l_isot_parts
      { % pop the entire match element (\0) from the front and discard it
        \seq_pop_left:NN \l_isot_parts \l_tmpa_tl
        \__isot_helper:ee {\seq_item:Nn \l_isot_parts 1} {\seq_item:Nn \l_isot_parts 2}
      }
      { #2 }
  }
}

\seq_new:N \l_rxnseq_parts
\regex_const:Nn \c_rxnseq_regex { ([(),\ ]) }

\NewDocumentCommand \rxnseq {m} {
  \regex_split:NnN \c_rxnseq_regex {#1} \l_rxnseq_parts
  \seq_remove_all:Nn \l_rxnseq_parts {}
  \seq_map_indexed_function:NN \l_rxnseq_parts \rxnseq_format_part:nn
}

\NewDocumentCommand \todo {s m} {
  {\color {ForestGreen} [
    \IfBooleanTF {#1} {#2} { TODO \tl_if_blank:nTF {#2} {} {:~#2} }
  ] }
}

\ExplSyntaxOff

\newcommand{\net}[1]{\texttt{#1}}
\newcommand{\cnoextras}{CNO\_extras}
\newcommand{\cnoheburn}{CNO\_He\_burn}
\newcommand{\enuc}{\ensuremath{\dot{e}_{\mathrm{nuc}}}}

\journalinfo{The Open Journal of Astrophysics}

\submitted{Version May 20, 2025}

\begin{document}

\title{Multi-dimensional Models of Mixed H/He Flames in X-ray Bursts}
\author{Eric T. Johnson\,\orcidlink{0000-0003-3603-6868} and Michael Zingale\,\orcidlink{0000-0001-8401-030X}}
\affiliation{Department of Physics and Astronomy, Stony Brook University, Stony Brook, NY 11794-3800, USA}

\begin{abstract}
We investigate the properties of mixed H/He flames in X-ray bursts using 2D hydrodynamic simulations.
We find that as the initial hydrogen abundance of the atmosphere increases, the flame is less energetic and propagates slower.
The simulation outcome, whether a flame forms and whether there's runaway burning at the base of the atmosphere, is very sensitive to the initial model and the nuclear reaction network used.
We also see that at late times a secondary flame ignites, with the ignition mechanism dependent on the composition.
\end{abstract}


\maketitle

%======================================================================
% Introduction
%======================================================================
\section{Introduction}\label{sec:introduction}

Type I X-ray bursts (XRBs) are transient outbursts from a low-mass X-ray binary system, triggered by thermonuclear burning on the surface of a neutron star \citep{galloway-keek}.
Depending on the accretion rate, the built-up fuel may be pure helium, or a mix of hydrogen and helium with varying proportions.
One-dimensional numerical simulations can accurately reproduce burst durations and recurrence times, light curves, and nucleosynthesis yields \citep{woosley-xrb,HEG_ETAL07,johnston:2018}.
However, some features such as burst oscillations and light curve rise times depend on the initial flame dynamics, which are inherently multidimensional \citep{watts:2012}.

As computational resources have become more powerful, more attention has been given to 2D and 3D simulations.
\cite{spitkovsky2002} used the shallow-water approximation to model flame propagation, and showed that the Coriolis force is important for flame dynamics.
However, this approximation does not model the vertical structure of the flame, so detailed investigations of flame dynamics and nucleosynthesis are not possible.
\cite{cavecchi:2013} and \cite{Cavecchi:2015} performed the first vertically resolved flame simulations, using a method that enforces vertical hydrostatic equilibrium.
These simulations used pure helium and showed an enhancement of the burning rate from the multi-dimensional flow pattern.
Follow-on full three-dimensional simulations find enhancement of the flame speed by a factor of 10 due to instabilities in the flame front \citep{cavecchi:2019}.
More recently, the formation of a hotspot, which may lead to the ignition of a burning front, was explored using 2D heat transport simulation by \citet{goodwin:2021}.

Previous work by our group has focused on pure helium flames, using the full compressible Euler equations modeled by the \castro{} simulation code \citep{castro,castro_joss}.
\cite{eiden:2020} introduced the simulation setup using a 2D axisymmetric domain, and investigated boosted flames by artificially increasing the reaction rates and thermal conductivity.
\cite{harpole:2021} looked at the effects of rotation rate and thermal structure, using more realistic simulations without boosting.
These calculations show the formation of a steady flame and explored how the flame ignition depends on the rotation rate.
These resolved simulations were enabled by porting our code to run on GPUs \citep{castro_gpu}.
These calculations also established the resolution requirements of ${\sim}\qty{20}{\cm\per zone}$.
\cite{xrb3d:2023} compared the 2D simulations to a full 3D simulation with the same parameters, and found that the two matched reasonably well, validating the use of axisymmetric geometry to explore the burning in the absence of turbulence.
Finally, \cite{chen:2023} explored the sensitivity of the flame to different reaction networks and found that including proton-capture reactions to bypass the slower \rxnseq{C12(a,g)O16} reaction caused a burst of acceleration at early times.

In this paper, we explore flames propagating through a mixed H/He atmosphere.
The addition of hydrogen enables addition nuclear pathways, creating a much more complex nucleosynthesis process than the pure He flames done previously.
In section~\ref{sec:numerical-methodology} we describe our numerical method, in section~\ref{sec:preliminary-simulations} we explore the size of the reaction network, in section~\ref{sec:results} we discuss our results, with an emphasis on the nucleosynthesis, and finally in section~\ref{sec:summary}, we conclude.

\section{Numerical Methodology}\label{sec:numerical-methodology}

All simulations are done with the \castro{} simulation code, which solves the compressible Euler equations together with thermal diffusion, reactions, and rotation (just the Coriolis force for this application).
The equations are supplemented with a general stellar equation of state from \cite{timmes_swesty:2000}.
We use the same initial model formulation and simulation parameters as \cite{xrb3d:2023}.
For reference, we use a peak temperature of \qty{1.4e9}{\K} in the hot region behind the flame and \qty{2e8}{\K} in the cool region ahead of the flame, a neutron star crust temperature of \qty{2e8}{\K}, a minimum temperature of \qty{8e6}{\K} at the top of the isentropic region, and a density of \qty{3.43e6}{\g\per\cm\cubed} at the interface between the neutron star and the atmosphere.
The main changes are the initial model composition and the nuclear reaction networks used.
The atmosphere is composed of uniformly mixed \isot{H1} and \isot{He4}, along with 1\% of \isot{O14} and \isot{O15} distributed according to their $\beta^+$-decay lifetimes to help start CNO burning.
Recently, \cite{guichandut:2024} explored a similar model starting from a layered H/He atmosphere, and found the convective flows could mix the composition efficiently.
Here, we assume that this mixing has already occurred before the flame ignition.

We use a two-dimensional axisymmetric domain of $\qty{196608}{\cm}\times\qty{24576}{\cm}$ with a coarse grid of $768\times192$ zones and 2 levels of refinement (first by a factor of 4 then by a factor of 2).
This gives a fine-grid resolution of \qty{32}{\cm} in the radial direction and \qty{16}{\cm} in the vertical direction (the same resolution as in \citealt{xrb3d:2023}).
The refined grids are static, with the finest resolution (level 2) present below $z=\qty{4096}{\cm}$ and the second finest (level 1) below $z=\qty{5120}{\cm}$.
This puts the initial atmosphere completely at the finest resolution.
We use a reflecting boundary at bottom of the domain, along with the well-balanced pressure reconstruction described in \cite{ppm_wellbalanced}.
At the top of the domain, we include an isothermal buffer region into which the flame can expand.
We also apply a sponge source term which suppresses the velocities in low-density regions \cite[described in][]{eiden:2020}, as we are not concerned with this material.

\begin{table}[t]
    \centering
    \small
    \caption{Overview of the reaction networks used}
    \label{tab:networks}
    \begin{tblr}{width=\linewidth, colspec={ c c c X[l] }}
    \toprule
        name & nuclei & rates & {\centering notes} \\
    \midrule
        \net{rprox} & 10 & 14 & based on Appendix C of \cite{wallacewoosley:1981}, more details in \cite{xrb2} \\
        \net{aprox19} & 19 & 78 & described in \cite{Kepler} \\
        \net{\cnoextras} & 21 & 67 & based on MESA \citep{mesa:2011}, only goes up to \isot{Mg24} \\
        \net{\cnoheburn} & 33 & 113 & \net{\cnoextras} + \net{subch\_simple} from \cite{chen:2023} \\
    \bottomrule
    \end{tblr}
\end{table}

\begin{figure*}
    \centering
    \includegraphics[width=\textwidth]{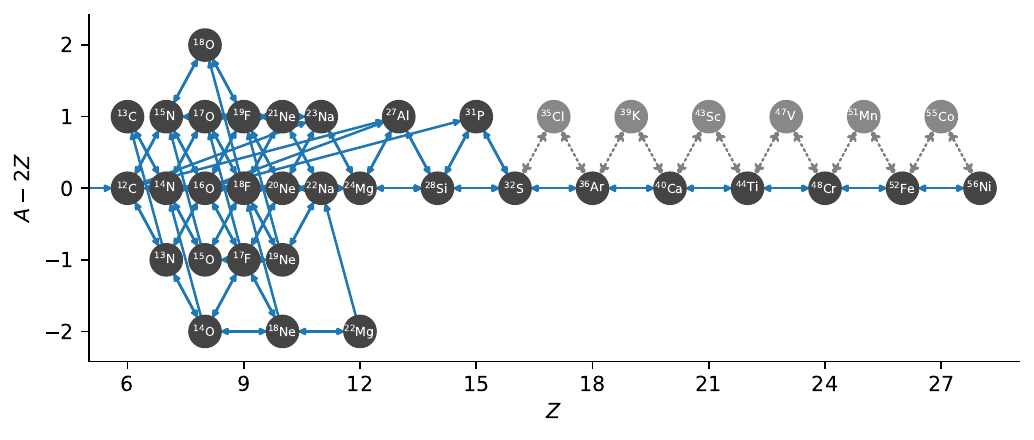}
    \caption[Structure of the \net{\cnoheburn} network]{Structure of the \net{\cnoheburn} network (links to \isot{H1} and \isot{He4} are not drawn). The light gray nodes and dotted lines represent the \rxnseq{(a,p)(p,g)} approximation, with those nuclei approximated out of the network.}
    \label{fig:network-plot}
\end{figure*}

We explore several different reaction networks, summarized in Table~\ref{tab:networks}, with the largest two (\net{\cnoextras} and \net{\cnoheburn}) created using \pynucastro~\citep{pynucastro2}.
All reaction rates come from JINA Reaclib \citep{reaclib}.
Reactions are coupled via Strang splitting, with the energy evolved together with the rate.
We use the \texttt{SCREEN5} routine for plasma screening from \cite{1982ApJ...258..696W}, neutrino cooling from \cite{itoh:1996}, and thermal conductivities from \cite{Timmes00}.

\section{Exploratory Simulations}\label{sec:preliminary-simulations}

Before creating \net{\cnoheburn}, we tried to create flames in our simulations using several other networks.
The burning front in these simulations would either fizzle out before setting up a propagating flame, and/or quickly ignite across the entire domain, similar to the behavior seen in the \texttt{F500\_4E8} run from \cite{harpole:2021}.
Our initial simulations with different networks are discussed in more detail in \cite{astronum-2023-eric}, but we give a brief description here.

The first set of runs we performed used \net{rprox}, a 10-isotope network introduced in \cite{xrb2} and based on the ``APPROX'' code from \cite{wallacewoosley:1981}.
This network includes approximations to hot CNO burning, triple-alpha, and rp-process breakout.
The simulations with this network did not show a propagating flame, but instead were disrupted by burning near the surface of the star, at the rightmost edge of the domain.
Using a lower temperature at the base of the fuel layer didn't help to prevent this burning.
Initial runs in this series also exposed a numerical issue with the hydrostatic boundary conditions used at the bottom of the domain, where material would flow into the domain and push up the rest of the atmosphere.
Switching to a reflecting boundary condition and well-balancing, as mentioned in section~\ref{sec:numerical-methodology}, resolved this by maintaining the hydrostatic equilibrium over long time scales.

We next ran a set of simulations using the \net{aprox19} network described in \cite{Kepler}.
We chose this network since it is widely-available, and commonly used in multi-dimensional simulations involving H burning.
As this network does not directly include \isot{O14} or \isot{O15}, we replaced these nuclei with \isot{C12} in the initial model.
These simulations were also unable to produce a stable flame, and had much more vigorous burning across the entire domain.
We note that this network was designed primarily for stellar evolution simulations, and does not include a full treatment of hot CNO burning, so the inability to generate a flame in these conditions is not surprising.

We had better results with \net{\cnoextras}, a 21-isotope network inspired by MESA \citep{mesa:2011} that includes hot CNO, breakout to \isot{Ne18} and \isot{Ne19}, and alpha-chain burning to \isot{Mg22} and \isot{Mg24}.
Using 9\% \isot{H1}, 90\% \isot{He4}, 0.35\% \isot{O14}, and 0.65\% \isot{O15}, we were able to get a stable flame propagating for \qty{2e4}{cm} before being disrupted by burning at the base ahead of the flame.
With this network, we were also able to decrease the peak temperature to \qty{5e7}{\kelvin} in the cool region, and still observe a flame which was sustained for longer, although it was significantly weaker.
This demonstrates the strong sensitivity of the flame properties on the thermodynamic state in the atmosphere.

Finally, we added additional alpha-chain nuclei up to \isot{Ni56} to \net{\cnoextras}, as well as \isot{ne21}, \isot{na22}, and \isot{na23} to create \net{\cnoheburn}, visualized in Figure~\ref{fig:network-plot}.
This produces robust flames.
The rest of this paper describes our results using this network.

\section{Effect of Initial Hydrogen Abundance}\label{sec:results}

We have performed five simulations using \net{\cnoheburn}, with varying initial abundances of hydrogen and helium: 0\% H, 5\% H, 10\% H, 20\% H, and 30\% H.
These were run out to at least \qty{150}{\ms}.
In the next subsections, we discuss the structure of the flame and the nucleosynthesis.

\subsection{Temperature and Energy Generation Evolution}\label{sec:weighted-profiles}

\begin{figure*}
    \centering
    \includegraphics{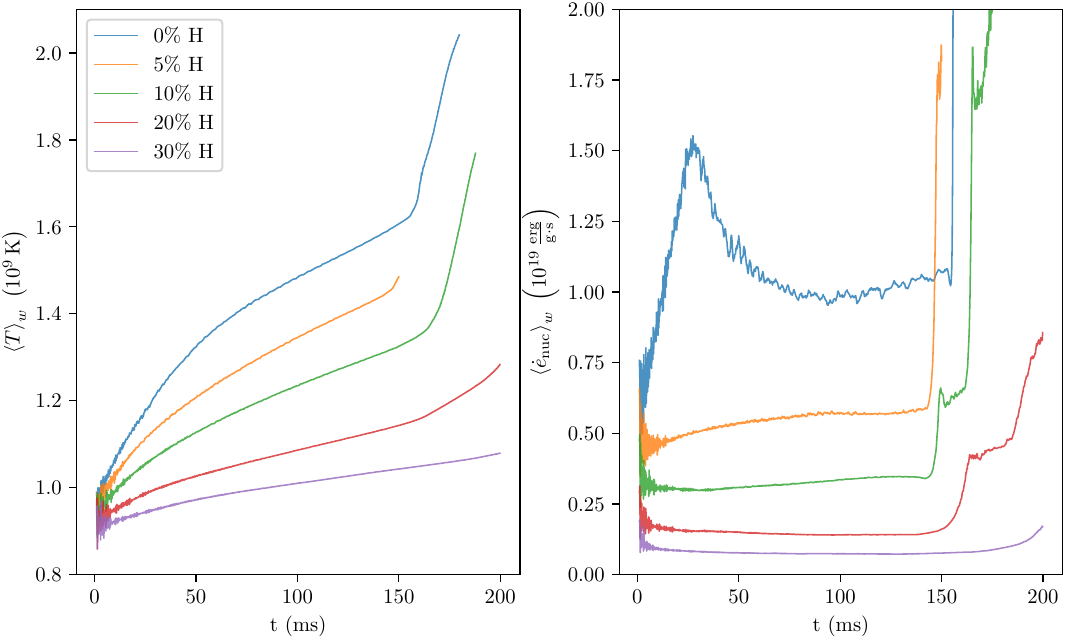}
    \caption[Temperature and energy generation rate in the burning region]{Weighted profiles of temperature (left panel) and energy generation rate (right panel) in the burning region for the five different initial compositions. In general, when more helium is present, the flame is hotter and the energy generation is higher.}
    \label{fig:weighted-profiles}
\end{figure*}

For a thermodynamic quantity $Q$, we define the density-weighted average value $\left<Q\right>_w$ as
\begin{equation}
\left<Q\right>_w = \frac{\sum_{c_i} \rho(c_i) Q(c_i)}{\sum_{c_i} \rho(c_i)}; c_i \in C_{99}(Q),
\end{equation}
where $C_{99}(Q)$ is the set of grid cells with $Q$ values in the top percentile over the entire domain, $Q(c_i)$ is the value of $Q$ in cell $c_i$, and $\rho(c_i)$ is the density in cell $c_i$.
This average allows us to track the thermodynamic properties of the whole flame more accurately than taking the maximum value, as the maximum is significantly noisier.
Figure~\ref{fig:weighted-profiles} shows the density-weighted average temperature and nuclear energy generation rate (\enuc) evolution over time, for each of the simulations.  While the overall evolutions are similar, there are a few differences between the pure He and mixed H/He runs.

The pure helium model shows similar trends to those seen in \cite{chen:2023}, with an initial spike in energy generation, although the present results are not as sharp or tall.
This could be an artifact of our initialization---if too large of a region is ignited at the star, we can burn too fast before a steady flame is established.
The abrupt peaks in \enuc{} after \qty{130}{\ms} are secondary flame ignitions at the base fuel layer, which we examine in more detail in section~\ref{sec:secondary-flame}.
As the fraction of hydrogen increases, the energy generation tends to be more constant over time, and the peak temperatures are lower.
However, all of the models eventually ignite a secondary flame.

\subsection{Flame Front}\label{sec:flame-front}

\begin{figure}[tp]
    \centering
    \includegraphics{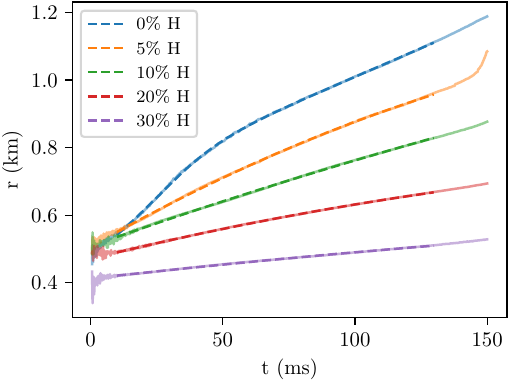}
    \caption[Flame front position over time]{Flame front position, tracked using 1\% of the global maximum of \enuc. Generally, the flame moves faster as the initial helium abundance increases.}
    \label{fig:flame-front}
\end{figure}

\begin{figure*}[tp]
    \centering
    \includegraphics{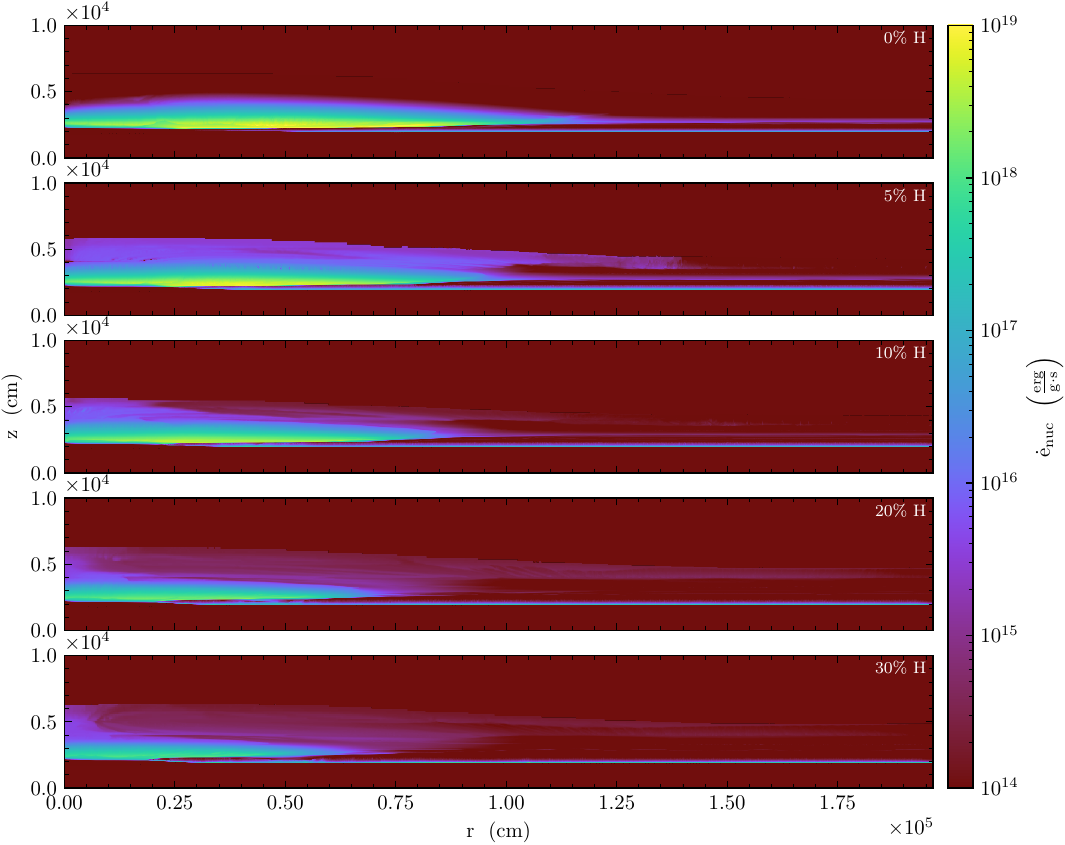}
    \caption[Nuclear energy generation at $t=\qty{100}{\ms}$]{Nuclear energy generation rate for the different initial hydrogen fractions at $t=\qty{100}{\ms}$.}
    \label{fig:comparison-100ms-enuc}
\end{figure*}

\begin{table*}[t]
    \centering
    \caption{Fit parameters for figure~\ref{fig:flame-front}}
    \label{tab:flame-speed-params}
    \begin{tabular}{lcccccc}
\toprule
    Run & $a_0$ (\unit{\km}) & $v_0$ (\unit{\km\per\s}) & $r_0$ (\unit{\km}) & $A$ (\unit{\km}) & $B$ (\unit{\ms}) & $C$ (\unit{\ms}) \\
\midrule
     0\% H & \num{2.7 +- 1.0} & \num{3.07 +- 0.10} & \num{0.5742 +- 0.0017} & \num{0.115 +- 0.004} & \num{24.2 +- 0.5} & \num{24.9 +- 0.3} \\
     5\% H & \num{-15.22 +- 0.13} & \num{4.466 +- 0.009} & \num{0.5061 +- 0.0003} & N/A & N/A & N/A \\
    10\% H & \num{-4.66 +- 0.13} & \num{2.775 +- 0.010} & \num{0.5081 +- 0.0003} & N/A & N/A & N/A \\
    20\% H & \num{-6.22 +- 0.06} & \num{1.919 +- 0.005} & \num{0.4710 +- 0.0001} & N/A & N/A & N/A \\
    30\% H & \num{-1.88 +- 0.05} & \num{0.872 +- 0.004} & \num{0.4124 +- 0.0001} & N/A & N/A & N/A \\
\bottomrule
\end{tabular}

\end{table*}

\begin{table}
    \centering
    \caption{Instantaneous flame propagation speed for the different runs}
    \label{tab:flame-inst-speed}
    \begin{tabular}{lccc}
\toprule
    Run & $v_{25}$ (\unit{\km\per\s}) & $v_{50}$ (\unit{\km\per\s}) & $v_{100}$ (\unit{\km\per\s}) \\
\midrule
     0\% H & 7.89 & 5.10 & 3.38 \\
     5\% H & 4.09 & 3.70 & 2.94 \\
    10\% H & 2.66 & 2.54 & 2.31 \\
    20\% H & 1.76 & 1.61 & 1.30 \\
    30\% H & 0.82 & 0.78 & 0.68 \\
\bottomrule
\end{tabular}

\end{table}

Figure~\ref{fig:flame-front} shows the position of the flame front over time for each simulation.
This is calculated by first vertically averaging the nuclear energy generation rate, limited to $z > \qty{2.2e4}{\cm}$ to exclude the burning at the base of the atmosphere.
Then, we find the first radial position after the local maximum where the value drops below 1\% of the global maximum (in both time and space). 
Fits are shown by the dashed lines.
We use a quadratic fit function for the hydrogen-enriched models, as in \cite{harpole:2021}:
\begin{equation}\label{eqn:quad-fit}
r(t) = \frac12 a_0 t^2 + v_0 t + r_0.
\end{equation}
For the pure helium model, we add a hyperbolic tangent term to equation~\ref{eqn:quad-fit} to 
capture the initial burst in acceleration, as in \cite{chen:2023}:
\begin{equation}\label{eqn:tanh-fit}
r(t) = A \tanh\left(\frac{t + C}{B}\right) + \frac12 a_0 t^2 + v_0 t + r_0,
\end{equation}
where $A$ is the amplitude of the acceleration burst, $B$ is proportional to the duration, and $C$ is the time of peak acceleration.
This acceleration corresponds to the peak in \enuc{} in figure~\ref{fig:weighted-profiles}.
All the fit parameters are shown in table~\ref{tab:flame-speed-params} and in table~\ref{tab:flame-inst-speed} we use the fits to evaluate the instantaneous speed at three different times ($v_{25}$, $v_{50}$, and $v_{100}$ are the flame speeds at 25, 50, and 100~ms respectively).
As table~\ref{tab:flame-inst-speed} shows, the flame propagation speed decreases as the initial hydrogen abundance is increased.
This is due to hydrogen burning being slower than helium burning, due to the beta-decay waiting points in the hot CNO cycle.
This also agrees with observational findings that mixed H/He bursts have longer timescales than pure helium bursts \citep{2017PASA...34...19G}.

Figure~\ref{fig:comparison-100ms-enuc} shows the flame structure at $t=\qty{100}{\ms}$.  We see that the lateral distance the burning front moves is inversely proportional to the amount of H in the atmosphere, consistent with the flame speeds computed above.
We also see that the peak energy generation rate decreases with H abundance, with the peak value in the 30\% H flame being 20 times smaller than that of the 0\% H flame.
Finally, we see that the overall geometry of the flame is similar in all cases, with the leading edge of the flame lifted off of the neutron star interface, as seen in \citet{cavecchi:2013} and \citet{eiden:2020}.

% get refinement boundaries: ds.index.grid_right_edge[ds.index.grid_levels[:, 0] == <level>][:, 1].max()
We note that the envelope of the pure helium flame reaches the jump from the finest level to the next coarsest level, above the atmosphere, at $t\approx\qty{30}{\ms}$ and the final jump to the coarsest level shortly thereafter at $t\approx\qty{55}{\ms}$.
The same occurs around \qty{65}{\ms} for the 5\% H run, \qty{80}{\ms} for the 10\% and 20\% H runs, and \qty{50}{\ms} for the 30\% H run.
Despite the fact that it has the weakest burning out of all the hydrogen-enriched simulations, the 30\% H flame reaches the refinement boundary before the other hydrogen-enriched flames.
This is due to the taller scale height of the atmosphere, which increases with the fraction of hydrogen present.
However, as there is not much burning at the top of the atmosphere, we do not expect the lower resolution there to greatly affect the flame physics.

\subsection{Nucleosynthesis}\label{sec:nucleosynthesis}

% 20.0 ms (run_0H/*_plt0087989), [ 53408,  2384] cm: rho=8.828099e+05 g/cm**3, T=5.472522e+08 K, enuc=1.963253e+18 erg/(g*s)
% 40.0 ms (run_0H/*_plt0174017), [ 53408,  2384] cm: rho=6.887832e+05 g/cm**3, T=8.514668e+08 K, enuc=6.036110e+18 erg/(g*s)
\begin{figure*}[tp]
    \centering
    \includegraphics{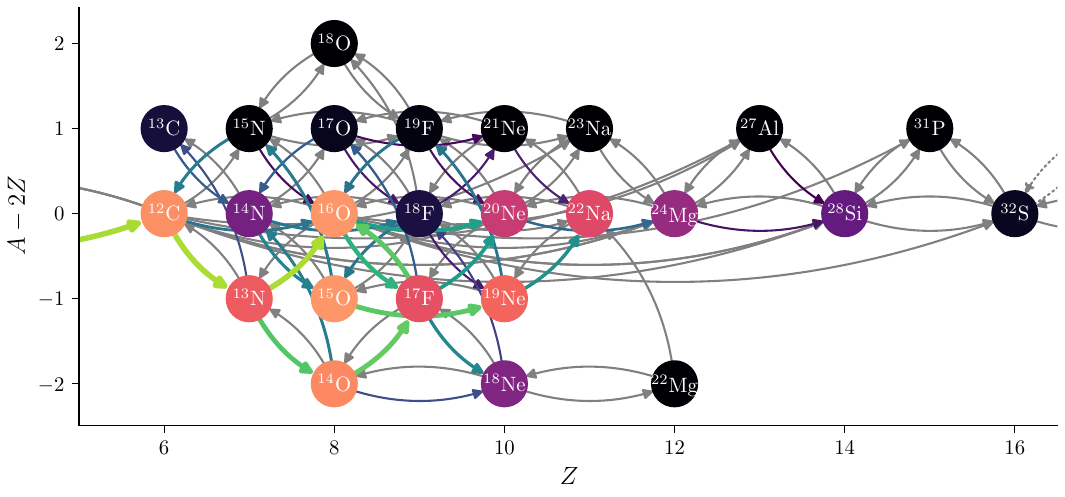}
    \includegraphics{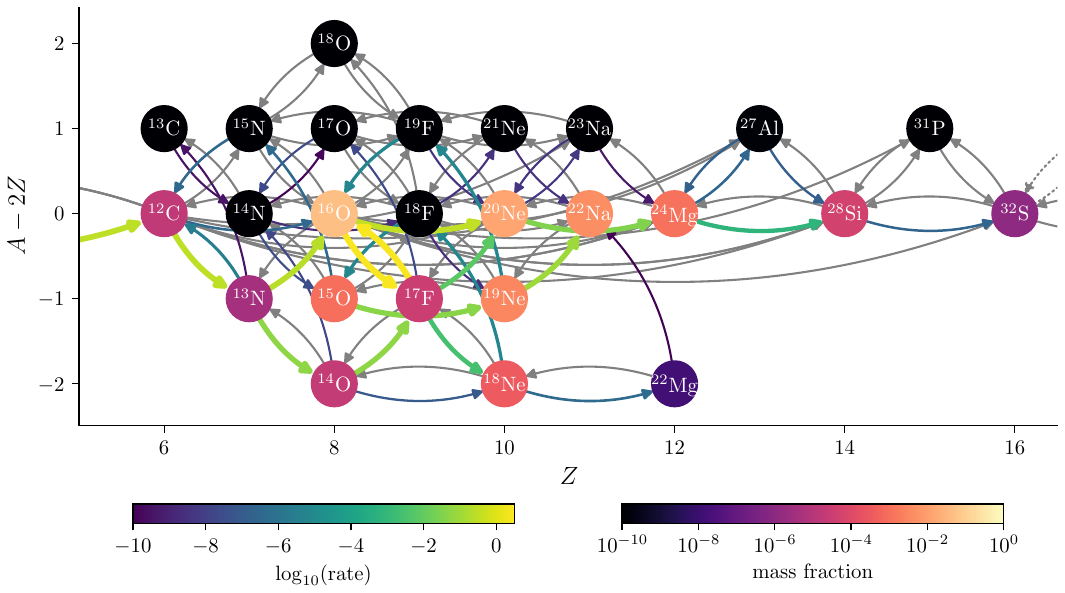}
    \caption[Reaction flow for 0\% H flame]{Snapshots of the reactions in the flame front (top panel, $t=\qty{20}{\ms}$) and behind the flame front (bottom panel, $t=\qty{40}{\ms}$) in the 0\% H simulation. Links are colored by the reaction rate strength and nodes are colored by the mass fraction of the corresponding species. The thermodynamic conditions are $\rho=\qty{8.83e+05}{\g\per\cm\cubed}$, $T=\qty{5.47e+08}{\K}$, $\enuc=\qty{1.96e+18}{\erg\per\g\per\s}$ for the top panel and $\rho=\qty{6.89e+05}{\g\per\cm\cubed}$, $T=\qty{8.51e+08}{\K}$, $\enuc=\qty{6.04e+18}{\erg\per\g\per\s}$ for the bottom panel.}
    \label{fig:reaction-flow-0H}
\end{figure*}

% 50.0 ms (run_10H/*_plt0217670), [ 54880,  2480] cm: rho=6.134689e+05 g/cm**3, T=6.488445e+08 K, enuc=2.410086e+18 erg/(g*s)
% 70.0 ms (run_10H/*_plt0303607), [ 54880,  2480] cm: rho=5.329466e+05 g/cm**3, T=8.196478e+08 K, enuc=2.345814e+18 erg/(g*s)
\begin{figure*}[tp]
    \centering
    \includegraphics{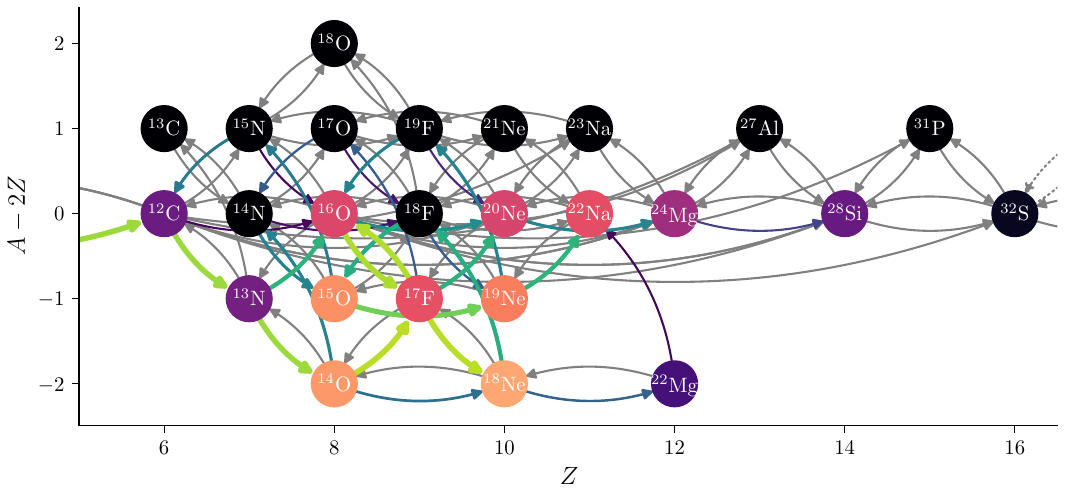}
    \includegraphics{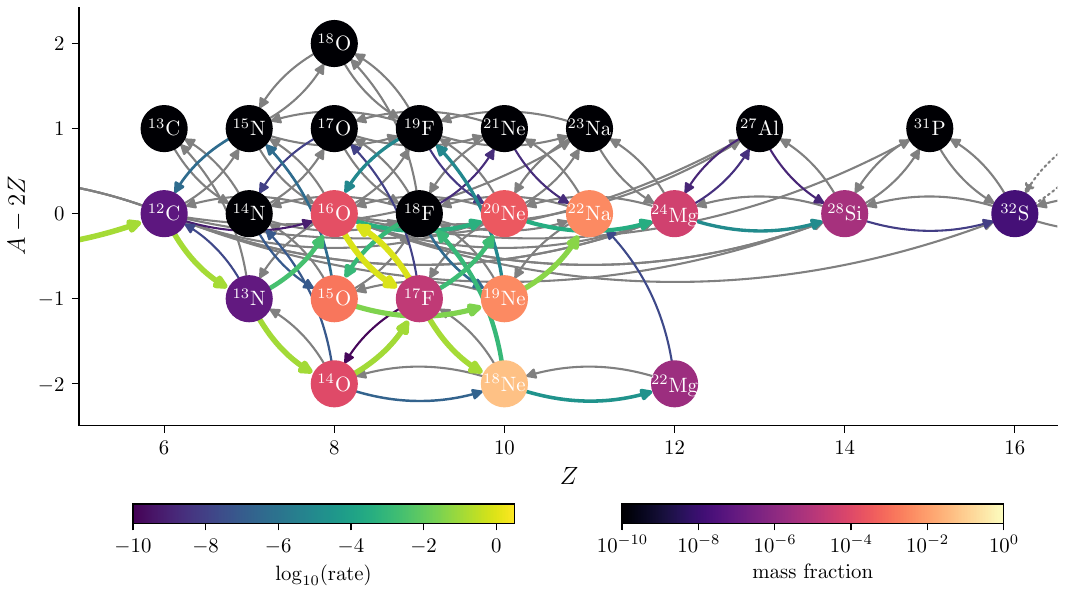}
    \caption[Reaction flow for 10\% H flame]{Snapshot of the reactions in the flame front (top panel, $t=\qty{50}{\ms}$) and behind the flame front (bottom panel, $t=\qty{70}{\ms}$) in the 10\% H simulation. Links are colored by the reaction rate strength and nodes are colored by the mass fraction of the corresponding species. The thermodynamic conditions are $\rho=\qty{6.13e+05}{\g\per\cm\cubed}$, $T=\qty{6.49e+08}{\K}$, $\enuc=\qty{2.41e+18}{\erg\per\g\per\s}$ for the top panel and $\rho=\qty{5.33e+05}{\g\per\cm\cubed}$, $T=\qty{8.20e+08}{\K}$, $\enuc=\qty{2.35e+18}{\erg\per\g\per\s}$ for the bottom panel.}
    \label{fig:reaction-flow-10H}
\end{figure*}

The use of \pynucastro{} to generate the reaction networks in \castro{} makes it fairly straightforward to investigate the specific reactions that are driving the flame, by allowing us to directly evaluate the reaction rates for any zone in our simulation.
Figure~\ref{fig:reaction-flow-0H} shows the abundances and reaction rate strengths for the pure helium case at two different times.
The main reaction sequence for this case is triple-alpha followed by \rxnseq{C12(p,g)N13(a,p)O16}, with this latter sequence identified in \citet{shenbildsten} as important for He burning for double detonation models of thermonuclear supernovae.
Our simulations, as well as those of \citet{chen:2023}, demonstrate that it is also important for X-ray bursts.
The body of the flame is hot enough for stable He burning beyond \isot{Mg24}, which is why this network performs better than \net{\cnoextras}.
However, underneath the flame, it is only hot enough for triple-alpha, leading to a build-up of a \isot{C12} layer.

For the hydrogen-rich models, figure~\ref{fig:reaction-flow-10H} shows the flow through the reaction network for typical conditions at the leading edge of the flame, as well as for those just behind the leading edge.
We see that proton capture onto \isot{N13} is preferred over beta-decay, as expected for hot CNO burning, and \isot{O14} builds up over time.
Inside the flame, \isot{O14} is mostly burned into \isot{Ne18} through \rxnseq{O14(a,p)F17(p,g)Ne18}.
Some of the intermediate \isot{F17} undergoes photodisintegration into \isot{O16}, which then proceeds to alpha-chain burning.
The complexity of this flow suggests that larger networks with more extensive rp-nuclei are needed, as explored in 1D in \cite{fisker:2008}.

\subsection{Secondary Flames}\label{sec:secondary-flame}

\begin{figure*}
    \centering
    \includegraphics{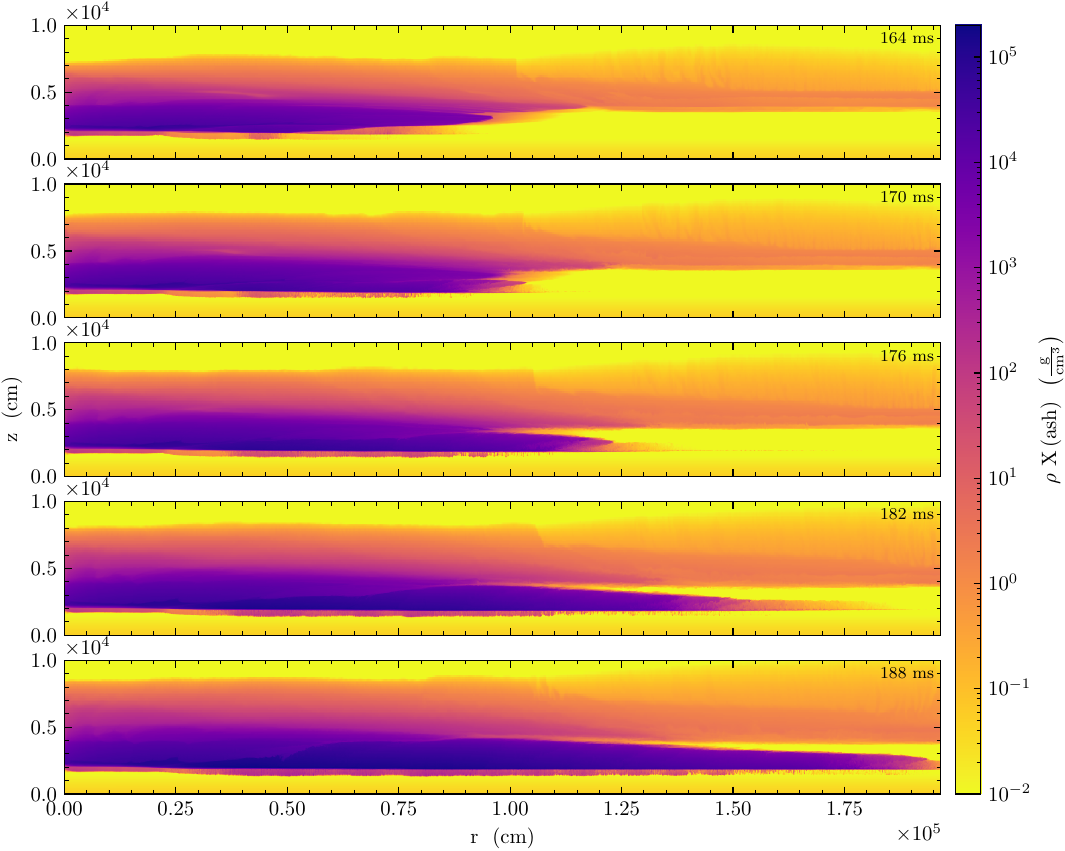}
    \caption[Time series of secondary flame for 10\% H]{Time series of the 10\% H run, showing the secondary flame that ignites at \qty{165}{\ms}. Here, ${\rm X}\left({\rm ash}\right)$ is the total mass fraction of isotopes not involved in any of the CNO cycles (sodium and beyond, as well as \isot{Ne20} and \isot{Ne21}), excluding \isot{Ni56} for better visibility against the NS crust.\vspace*{0pt}}
    \label{fig:second-flame-10H}
\end{figure*}

A new feature in these simulations is the ``secondary flame'' at late times.
Previous simulations by our group did not run out as far in time as the present results, and thus did not see this behavior.
We observe a very fast (around \qty{30}{\km\per\s}), strong secondary deflagration near the base of the atmosphere starting at \qty{160}{\ms} for 0\%~H, \qty{148}{\ms} for 5\%~H (visible in figure~\ref{fig:flame-front}), \qty{165}{\ms} for 10\%~H, and \qty{195}{\ms} for 20\%~H.
The time evolution of the flame for the 10\%~H run is shown in figure~\ref{fig:second-flame-10H}.

% ds.point((8.69120e+04, 2.24000e+03, 0))
For the pure helium run, this is due to the build-up of \isot{C12} over time at the base of the atmosphere from triple-alpha.
Once this layer gets hot enough ($\approx\qty{5e8}{\K}$), the proton-capture reaction \rxnseq{C12(p,g)N13(a,p)O16} bypasses the slower \rxnseq{C12(a,g)O16} reaction \citep{weinberg:2006,chen:2023}.
This allows further alpha-chain burning, creating a positive feedback loop which quickly consumes all of the \isot{C12}.

The mechanism for the simulations with hydrogen works similarly, but with a different set of reactions.
% ds.point((6.87360e+04, 2.36800e+03, 0))
At the base of the atmosphere, underneath the flame, the temperature gradually increases due to hot CNO and triple-alpha burning.
The triple-alpha also creates an excess of \isot{O14} via \rxnseq{C12(p,g)N13(p,g)O14}.
Once the temperature there reaches $\sim$\qty{7e8}{\K}, \rxnseq{O14(a,p)F17} goes faster than the beta-decay reaction, releasing a large amount of energy (a total of \qty{5.11}{MeV/nucleon} when including the faster \rxnseq{F17(p,g)Ne18}) and driving the secondary flame.

It remains a topic of future investigation as to whether this same behavior is seen with different initial models (especially the base density and temperature), or with different initialization methods.

\section{Summary}\label{sec:summary}

We perform the first multidimensional simulations of mixed H/He flames in X-ray bursts.
We observe that the more hydrogen present in the initial model, the slower the flame propagates, consistent with
weak-rate waiting points.
We also find that relatively large reaction networks are needed to produce these flames.
We explore the reaction mechanisms responsible for driving the flame, and identify the different reaction flows between a pure helium burst and mixed H/He bursts.
Finally, we observe a secondary flame ignition that requires further study.

Follow-on simulations will focus on modeling the entire star from pole to pole, which allows us to more realistically evolve for longer time scales and explore the effects of the changing Coriolis force at different latitudes.
We will also explore different initial models, as the flame is very sensitive to burning at the base of the atmosphere.
Finally, we will also consider even larger networks, with more rp-process nuclei, to better understand the nucleosynthesis.

\section*{Acknowledgments}

We thank S.\ Guichandut for helpful discussions on flame ignition.
\castro{} and the AMReX-Astrophysics suite are freely available at \url{http://github.com/AMReX-Astro/}.
All of the metadata and global diagnostics are available in a Zenodo record as \cite{h_he_xrb_metadata}.
The work at Stony Brook was supported by DOE/Office of Nuclear Physics grant DE-FG02-87ER40317.
This research used resources of the National Energy Research Scientific Computing Center (NERSC), a Department of Energy Office of Science User Facility using NERSC award NP-ERCAP0027167.
This research used resources of the Oak Ridge Leadership Computing Facility at the Oak Ridge National Laboratory, which is supported by the Office of Science of the U.S.\ Department of Energy under Contract No.\ DE-AC05-00OR22725.

{\vskip6pt{\it Software:}
    \amrex{} \citep{amrex_joss},
    \castro{} \citep{castro, castro_joss},
    GNU Compiler Collection (\url{https://gcc.gnu.org/}),
    GNU Parallel \citep{tange_2024_14550073},
    Linux (\url{https://www.kernel.org}),
    matplotlib (\citealt{Hunter:2007},  \url{http://matplotlib.org/}),
    NumPy \citep{numpy,numpy2,numpy2020},
    \pynucastro{} \citep{pynucastro,pynucastro2},
    Python (\url{https://www.python.org/}),
    SciPy \citep{scipy,scipy2},
    SymPy \citep{sympy},
    valgrind \citep{valgrind},
    yt \citep{yt}
}

{\vskip6pt{\it Facilities:}
    NERSC, OLCF
}

\bibliographystyle{aasjournal}

% Final typesetting pass: this is needed to prevent the caption for figure \ref{fig:comparison-100ms-enuc} from overlapping the text
\clearpage

% You should give the same name for your .bbl as your main .tex
% since it is a requirement for posting on ArXiv.
\bibliography{h_he_xrb}

\end{document}